# Procedural Low-Poly Terrain Generation

Landscape Generation with Terracing for Low-Poly Computer Games

Richard Tivolt
Faculty of Engineering
University of Southern Denmark
richard.tivolt@gmail.com

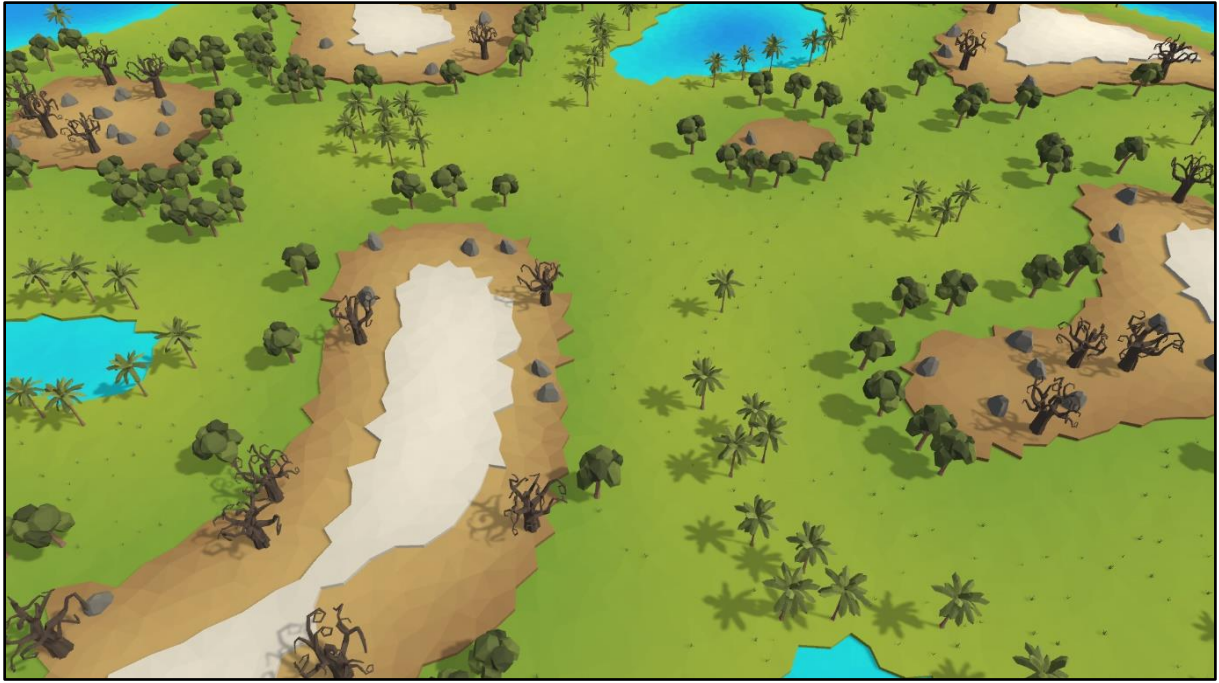

**Figure 1:** *Example of a terrain generated by the proposed technique and algorithms in this paper.*

## Abstract

In computer games, traditional procedural terrain generation relies on a grid of vertices, with each point representing terrain elevation. For each square in the grid, two triangles are created by connecting fixed vertex indices, resulting in a continuous 3D surface. While this method is efficient for modelling smooth terrain, the grid-like structure lacks the distinct, chaotic appearance of low-poly objects and is not suitable to be used for our purposes.

The technique presented in this paper aims to solve the following problem: Generate random, low-poly looking terraced terrain with different biomes and add vegetation to create an interesting environment.

## Introduction

Procedural terrain generation has become a highly viable technique among game developers, due to its ability to make infinite amount of content and gameplay for the player. Games like Minecraft, Terraria or No Man's Sky are fun to play, because every time a world is created, it is unique and different from all the other. This variability allows players to enjoy the game without the experience becoming repetitive or predictable.

By providing different landscapes, resources and challenges in each playthrough, players are kept engaged and encouraged to explore the entire world. As a result, games using procedural terrain generation can keep players invested for longer periods of time, creating an overall better gaming experience.

For smaller indie companies, procedural terrain generation is a great tool to avoid large amount of level and map design. Creating a proper level generator can highly increase playtime and variability, while still producing game content with sufficient quality.

## Related Work

Procedural techniques began to develop in the late 20th century. Early works included the exploration of fractals and mathematical functions for visual representation.

The introduction of fractal geometry in the 1980s, particularly through Benoît Mandelbrot's seminal work in his 1982 book, *The Fractal Geometry of Nature*, popularized the use of recursive algorithms to create complex, self-similar patterns.

In 1983, Ken Perlin introduced Perlin noise, a gradient noise function that became a cornerstone for generating natural-looking textures and terrains. This innovation allowed for the creation of coherent randomness, which is essential in procedural content generation.

The 1990s saw the practical application of procedural techniques in video games. Notable examples include Elite (1984), which generated a universe of star systems, and Doom (1993), which used procedural techniques for level design. The use of heightmaps for terrain generation also gained popularity in this time, allowing developers to create random looking 3D landscapes for games and even movies, like for the movie Tron (1982).

In the 2000s, advancements in computing power unlocked the possibility of real-time procedural techniques. Methods such as tile-based generation, Voronoi diagrams and Delaunay triangulation became commonly used in game development, allowing the creation of more complex environments.

Recently, in the 2010s and beyond, hybrid techniques have gained popularity by combining multiple procedural methods, resulting in even more complicated environments. Games like Minecraft, Terraria and No Man's Sky are notable examples of what can be achieved with the power of procedural terrain generation.

## Methodology

This section of the paper will go over the key steps of procedural low-poly terrain generation, and give insight on how to generate a terrain, like the one showcased in Figure 1.

### Generating Points

The main difference between traditional and low-poly terrains is the way mesh vertices are arranged. Instead of a fixed structure, our objective is to generate random points on a plane and connect them into triangles. This task is trivial using a grid, because the position and neighbours of the vertices are predictable, however it proves to be much more difficult, when the points are randomly placed.

Our first instinct would be to start generating random X and Y coordinates for our points in a given boundary. On a small scale this might satisfy our needs, but as shown in Figure 2, most of the time the points will be cluttered up or have too much space between them, making the overall shape of the terrain look uneven.

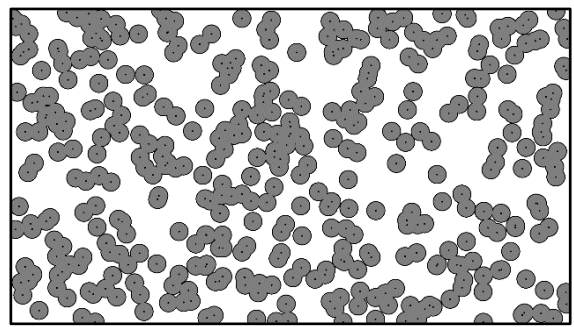

**Figure 2:** *Set of randomly distributed points.*

The problem arises from the fact that we do not account for the points being too close to one another. An approach to solve this would be to assign a minimum distance radius to the algorithm and check if a point is far enough from all the others, before adding it to our collection. Unfortunately, this method has a time complexity of $O(n^2)$, which becomes very slow when generating a large set of points. Moreover, as shown in Figure 3, the algorithm does not make use of the available space as much as possible, lowering the overall quality of our terrain.

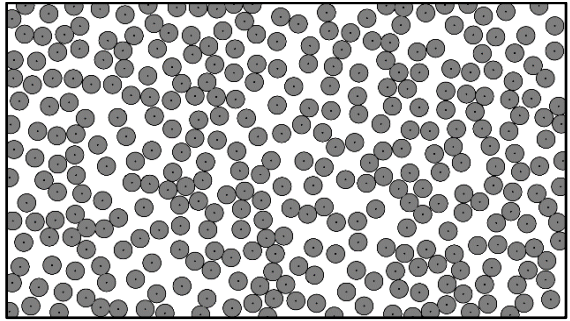

**Figure 3:** *Set of randomly distributed points, accounting for minimum distance.*

An optimal solution would be using an efficient uniform distribution algorithm, like Poisson-Disc Sampling, which ensures that none of the points are overlapping, while packing them tightly in $O(n)$ time.

The algorithm works by creating a grid, where the diagonal of a cell is the same size as the radius of our points. This ensures that in any given cell, there can only be one point, because if there were more, they would overlap. First it picks a random point inside the sample region, and places it into its corresponding cell. Then the algorithm will iteratively try to create more points around the current spawning point, while recording them as new available spawning candidates. When a point is being added, it only has to check the neighbouring five-by-five grid for any possible overlap. This will continue until there are no more spawn points left, and the entire sample region is filled. As shown in Figure 4, this will create a much better distribution than our previous attempts.

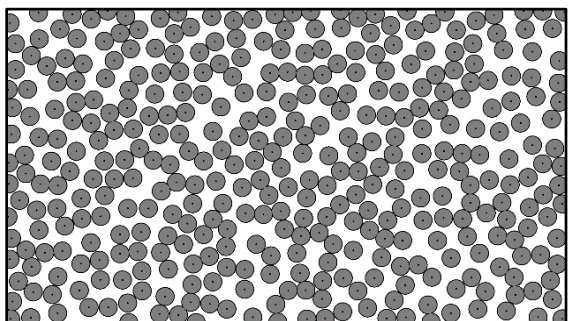

**Figure 4:** *Set of points generated by the Poisson-Disc Sampling algorithm.*

**Expanding to Square Shape**

The following step is not required to be included in the process, but it is recommended, as it makes hiding the edges of the terrain from the player much easier.

Since all the points are within the bounding box of the sample region, we can generate evenly spaced points around its circumference. We can do this multiple times, while increasing the distance from the previous iteration. As shown in Figure 5, at the end of this cycle the terrain will be surrounded by a large area, which can be used to hide the edges of the map, for instance, by making it into a large area of water.

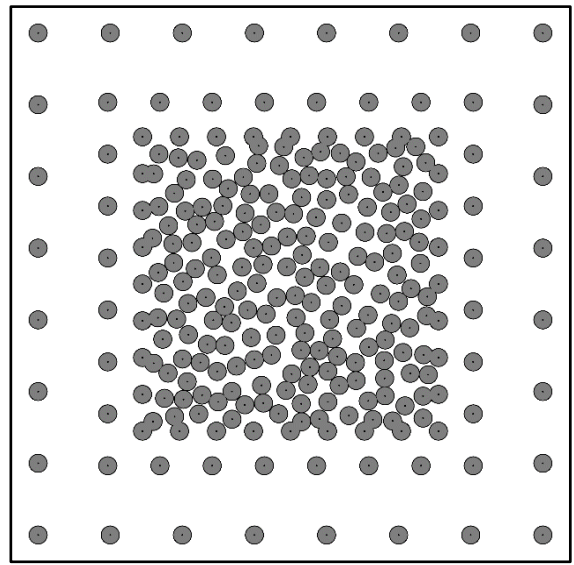

**Figure 5:** *Set of random points expanded into a square shape.*

Without making any checks while expanding the terrain, some of the extension points might get too close to the original ones. This can be corrected, but the difference it makes is subtle, so for our purposes, this implementation will be sufficient.

**Triangulating Points**

Now that we have our set of points, we have to find a way to connect them into triangles. In a grid-like structure, calculating the neighbours of each point is straight forward, because the points are arranged in a predictable pattern, making it easy to find adjacent points based on their indices. With their position randomized this predictability is lost, and determining the neighbouring vertices becomes a much more complicated task.

One approach to solve this problem is generating a Delaunay Triangulation from our points. This technique creates a set of triangles, where no point lies inside the circumcircle of any triangle in the set.

To compute this triangulation, we will use a technique called the Bowyer-Watson algorithm, which can generate a Delaunay Triangulation in $O(n\ log n)$ time. First, it constructs a triangle that contains all the points, known as the "super-triangle", and adds it to the set. Then, it picks a random point and finds all the triangles whose circumcircles contain that point. For the first iteration, this will only include the super-triangle. Next, these triangles are removed from the set, and their vertices are connected to the current point, forming new triangles, which are then added back to the set. After repeating this step for all the points, the triangles sharing the vertices of the super-triangle are removed, and the triangulation is completed.

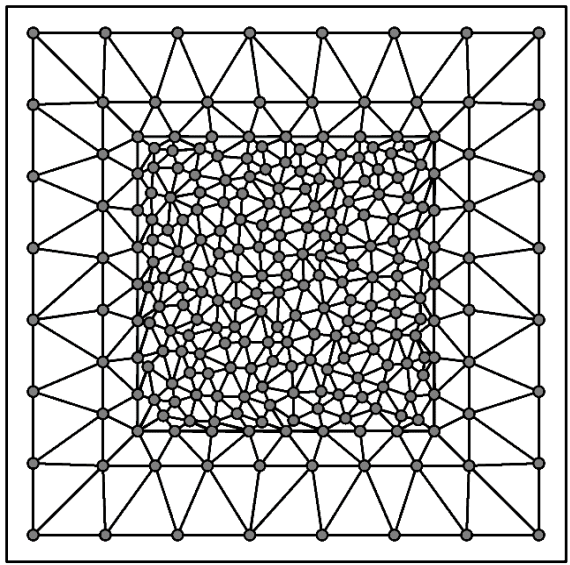

**Figure 6:** *Delaunay Triangulation of a set of points.*

Using this method, we can connect the points together and form a continuous low-poly surface. This will serve as the base mesh for the terrain, which can later be rendered by any 3D game engine.

**Generating Heightmap**

To bring our set of flat points into the third dimension, we need to find a way to assign a height value to each point, so that after elevating them, we get a smooth-changing terrain.

A handy function for this is Perlin noise, a gradient function that generates smoothly changing values between 0 and 1 for any given input. After running the points through the noise function, our surface gets transformed into a simple-looking terrain with hills and valleys.

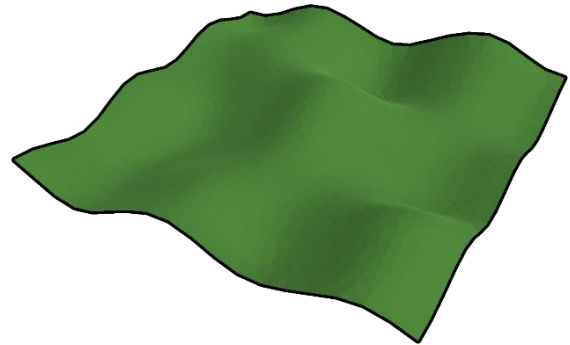

**Figure 7:** *Terrain vertices being elevated by the Perlin noise function.*

A technique called Noise Layering can be used to further improve our terrain, by adding multiple layers of noise on top of each other with increasing frequency and decreasing amplitude.

This will make the terrain look more detailed and interesting but also allow the height values to exceed the 0 to 1 boundary. Due to this reason, we will have to keep track of the lowest and highest values while generating them and map the end results back to the 0 to 1 range.

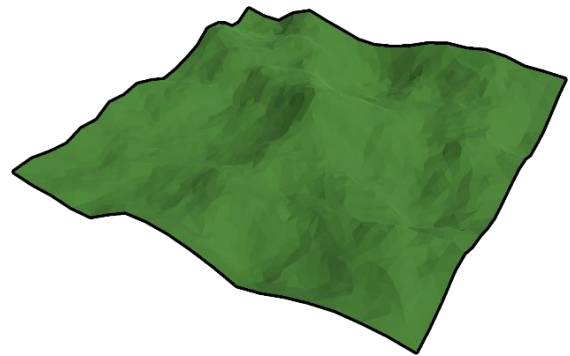

**Figure 8:** *Detailed terrain using Noise Layering.*

The current state of our heightmap is adequate for generating the terrain and can be used for creating the map. However, making the map closed, like an island, can be beneficial for limiting the available space and resources the player has access to.

A falloff map can be applied to the heightmap to gradually lower the edges of the terrain to a ground level. For our purposes, a simple linear function will suffice. By applying the function to the height values, those above the falloff boundary remain unchanged, while the ones within the boundary are interpolated between their original height and the ground level. Any values below the boundary are set to the given ground level. In this paper, the boundary is determined by the distance between the points and the edges of the map.

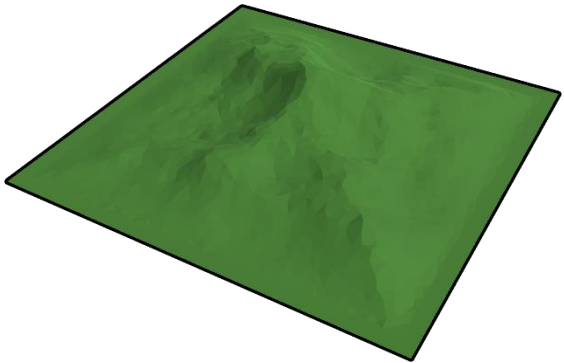

**Figure 9:** *Terrain after applying falloff map.*

Finally, using the heightmap the vertices of the triangulated mesh can be elevated, and the mesh can be rendered by a game engine.

**Forming Terraces**

What makes the technique presented by the paper unique is the fact, that the terrain has terraces instead of its height changing in a continuous manner.

To achieve this appearance, the height values are divided into distinct height bands, also referred to as biomes. Since the noise values are between 0 and 1, distinct biomes can be defined by a lower and upper boundary within this range. The points are then assigned to these biomes, by checking which boundary does their noise value fall into. After that, by setting all the height values in a biome to the same value, the terrain gets divided into distinct regions.

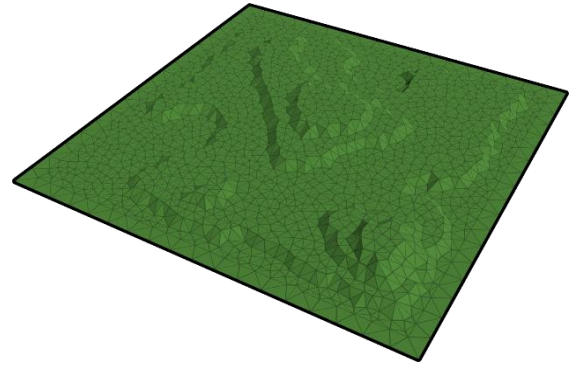

**Figure 10:** *Vertices divided into distinct bands.*

After flattening the biomes, there will be some rising triangles left at the borders of the regions. Identifying these is an easy task, because out of their three vertices, one of them will be assigned to a different biome. To form the terraces, all vertices in the upper biome will be moved to the lower one, making the terrain completely flat.

In some edge cases an issue arises from the fact that triangles can cover more than two biomes. Since the noise values are unpredictable, it is possible that a triangle will have all three of its vertices in different biomes or have them more than one biome "away" from each other. These triangles are considered invalid, because when this happens, holes will show up in the mesh. Fortunately, this can be easily detected, by checking the difference between the biomes of the triangle's vertices.

An easy fix to this problem is using a different seed for generating the points or for generating the noise values. A better but more complicated solution is to move the triangle vertices towards their lowest biome, until all the triangles become valid, and there are no more holes left.

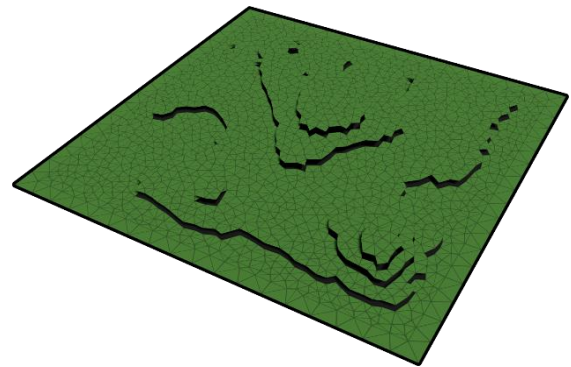

**Figure 11:** *Rising triangles flattened to the ground.*

There are two types of rising triangles: one with a single vertex, and another with two vertices in the upper biome, the former being used for the next step. Using the two vertices that were moved to the lower biome, two more vertices can be calculated above them in the upper biome. These are then connected to form the walls and complete the mesh of the terrain.

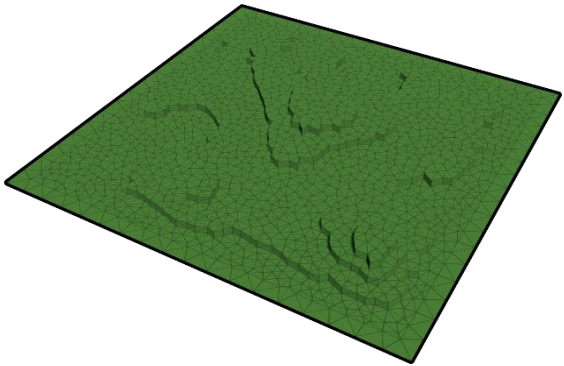

**Figure 12:** *Final terrain mesh with terraced walls.*

In some special cases, two triangles with two of their vertices being in the upper biome will share these two vertices. In this case, both of them will be moved to the lower biome, but a wall will be formed at their shared edge. When this happens, the wall is marked as invalid, and removed from the mesh, because it is not connected to any other surface.

**Colouring Triangles**

Now that the mesh of the terrain is completed, we can start colouring it. Since the triangles are divided into biomes, a gradient can be assigned to each region and used to interpolate between colours based on the average noise value of their vertices. This will make triangles close to each other look similar and allow us to add features like shallow and deep water or different kinds of grass.

Another colour should be assigned and used to colour the walls. Several methods can be used to determine, whether a triangle is part of a wall or not, but the easiest one is checking what direction does its normal vector points to and calculate its dot product with another vector pointing upwards. If this product is 0, the triangle is part of a wall.

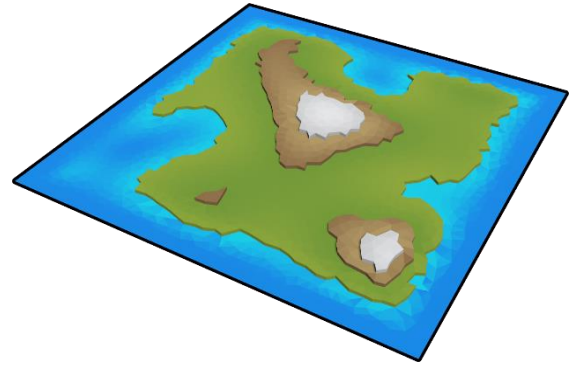

**Figure 13:** *Biomes coloured based on noise value.*

An additional step can be added to further improve the quality of the terrain, by offsetting the height of each vertex in the mesh. This will make the normal vectors of the triangles point to slightly different directions, allowing the lighting of the render engine to break up large patches of similar colours.

In case it is important to keep the surface of the regions flat, changing the noise value instead of the height value before calculating the colours of the triangles will achieve similar results. However, this can affect vegetation generation, so it is highly advised to set the noise values back, after the triangles are recoloured.

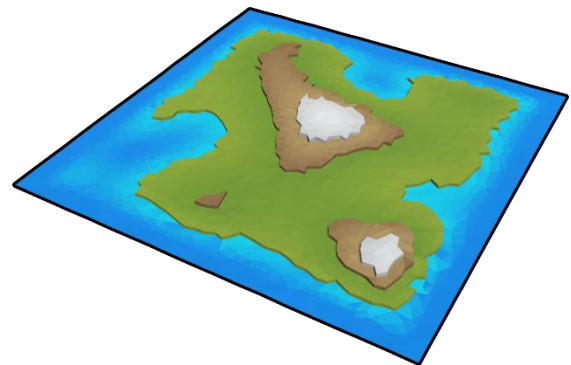

**Figure 14:** *Adding random height offset to vertices.*

**Adding Environment**

The technical portion of the paper is now complete; we have finished the terrain generation algorithm. However, to make the map more engaging and playable, additional elements like trees and rocks can be added for the player to interact with. There are various methods to place objects on the terrain, but one particular approach makes great use of the previously implemented algorithms.

By modifying the triangulation script to loop through the triangles, and store the connections between vertices, the entire mesh can be represented as a graph. This can be used to select random vertices, and place objects around them by visiting their neighbours, forming patches of resources like ores or dense forests.

A different technique utilizes the noise values of the vertices. Similar to the biomes, height bands can be assigned to vegetations and placed according to each vertex's noise value. As shown in the example, this can be used to form tree lines around the shore, or to keep certain areas clear for the player to spawn.

To prevent the objects from overlapping, the graph representation can be used to mark neighbouring vertices invalid after placing an object. This will ensure, that no object is positioned too close to another one and allow us to move them by some random offset, making the environment look even more natural.

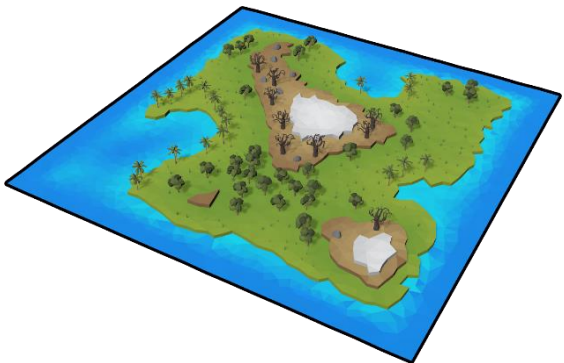

**Figure 15:** *Final map after adding environment.*

## Implementation

The implementation of the proposed terrain generation method was carried out in Unity using C#. The development process involved transforming the theoretical steps discussed in Methodology into Unity's environment, making use of its built-in classes and mesh rendering features.

Scripts were organized to follow a modular design, allowing each core functionality to operate independently. This approach ensures that each part of the algorithm can be modified or optimized individually without affecting other components. The implementation of the algorithm is available on GitHub: [link]

### Generating Mesh Data

The primary terrain generation script begins by running the Poisson Disc Sampling algorithm to generate the initial vertices for the mesh and store them in a list of Vector3s. After that, the points are expanded to a square shape, by defining the four vertices of a square around the sample region, linearly interpolating between them, and adding the points to the list.

The vertices are then fed into the Delaunay Triangulation algorithm, and the triangles are stored in a list. Since Unity's Vector3 class is a struct, the points are converted into our own Vertex class, so triangles sharing common vertices can reference the same object.

For later usage, the triangles and their vertices are stored in two separate lists, where each vertex only appears once. To create the graph representation of the terrain, we can iterate over the list of triangles and add their vertices to each other as neighbours.

### Applying Heightmap

Unity provides a built-in Perlin noise function in its Mathf class, but it only works for 2D points, so the vertices are converted into Vector2s using their X and Z coordinates. After running all the points through the noise function, the values are stored in an array of floats and normalized to a 0 to 1 range using the lowest and highest noise value.

To apply the falloff map we can iterate over the vertices, check their distance from the edges, and linearly interpolate between their noise value and 0 based on that distance. After that, the noise values are assigned to the vertices, and their biomes are calculated.

The final step is to ensure that each triangle spans at most two biomes. This can be achieved by iterating as many times as there are available biomes. During each iteration, all triangles are checked to find the lowest and highest biomes among their vertices. If the difference between the biomes is more than one, vertices not in the lowest biome are lowered to the next biome above the lowest one. After the biomes are settled, the Y coordinates of the vertices are calculated by multiplying the biome with a terrace height value. The noise values are recalculated by averaging the noise values of neighbouring vertices, while keeping the result between the lowest and highest possible values for that particular biome.

**Creating Triangles**

As it was previously explained in Methodology: "In some special cases, two triangles with two of their vertices being in the upper biome will share these two vertices. In this case, both of them will be moved to the lower biome, but a wall will be formed at their common edge."

A solution for this unwanted behaviour, is to have the triangles store their validity for being used in creating walls. First, we make a dictionary for associating a pair of vertices with a triangle. Then we loop over all the triangles in the list created in the first step, and check if they have their vertices in different biomes. If they do, and two of them are in the upper one, we add those two vertices with the triangle to the dictionary. If the dictionary already has a different triangle associated with those two vertices, we mark both triangles as invalid.

Now that everything is ready, we can start iterating over all the triangles, and adding them to the mesh, based on the following rules; if all three of the triangle's vertices are in the same biome, add them to the mesh, and assign them a colour based on their average noise value. If the vertices are in different biomes, the triangle is a rising triangle, and it will need to be transformed before adding it to the mesh.

Due to the fact, that triangles span at most two biomes, it is guaranteed, that one of the vertices – referred to as the "leading vertex" – will lie in either the upper or lower biome, while the other two will lie in the opposite one. If the leading vertex is in the upper biome, the triangle is flattened by lowering the leading vertex to the level of the other two. After that, the triangle can be added to the mesh the same way as before. If the leading vertex is in the lower biome, the other two vertices are lowered the same way as for the other case. Next, if the current triangle is valid for creating a wall, two more triangles are constructed using the following vertices: the two vertices of the current triangle that were in the upper biome, along with those same two vertices elevated back to the upper biome. With these four vertices we can construct two more triangles and fill the holes between the biomes.

**Finishing Terrain**

As a final touch, the mesh vertices are moved around a bit by a random offset to add more detail to the terrain. Since the mesh contains duplicate vertices to achieve flat shading, it is important to store these offsets in a dictionary to ensure that duplicate vertices are moved by the same amount.

Using Unity's built-in Mesh class, the vertices, triangles, and colour values are assigned to a new mesh, along with placeholder UV coordinates. After recalculating the face normals and bounds using the built-in functions, the mesh is assigned to a MeshFilter and rendered using a MeshRenderer. Additionally, a MeshCollider can be added to allow physics objects to interact with the terrain.

**Generating Environment**

Bringing our plain terrain to life can significantly improve its final look. Several methods can be used to determine spawning points for objects, like ray casting or selecting random points and checking for collisions. However, there is a much more efficient way, made possible by the way the terrain generation was implemented.

The algorithm is the following: for all terrain vertices, first check if the vertex lies on the edge of a biome. This can be determined by checking the biome of neighbouring vertices. If it does, skip that vertex, otherwise, based on the selected vertex's noise or biome value, spawn an object there. The Poisson Disc Sampling algorithm guarantees that no two vertices are closer to each other than a given distance. This fact, along with the graph-like structure of the terrain vertices is used to mark neighbouring vertices as invalid for spawning new objects. Depending on the distance between vertices and the size of the initial object, deeper neighbouring vertices can also be marked to prevent overlap using a simple graph traversal algorithm like depth-first search or breadth-first search.

Spawning objects for every vertex can lead to an excessive number of entities. To prevent this, a spawn probability is assigned to each object type. This allows for more control over the density of the environment and the number of objects in the scene.

## Results

The terrains are highly customizable, which allows this algorithm to be used in a lot of game genres. There are several parameters that can be tuned to customize the style of the terrain for different game scenarios, but it is recommended to use it for scenes with a top-down view. The following examples were generated by changing the vertex count, noise settings, biomes, environment settings, detail offset and disabling certain parts of the algorithm.

By decreasing the minimum vertex distance or increasing the sample region, the Poisson Disc Sampling algorithm is able to generate much more vertices and thus detail to the terrain. By lowering the noise scale and adjusting the probability values for the environment, we can generate much bigger maps.

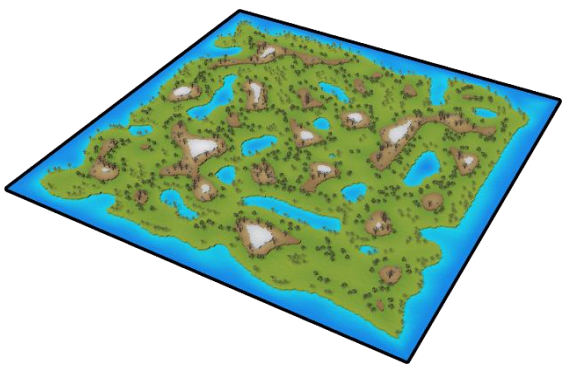

**Figure 16:** *A large map generated for bigger gameplay scenarios.*

To vary the theme and ambience of the terrain, we can change the biome colours. This can be used to introduce new worlds to the player, which is great for further expanding the contents of the game.

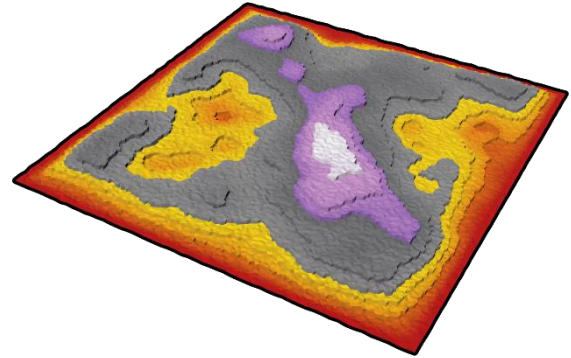

**Figure 17:** *A "Hell" themed map made by changing the biome colours.*

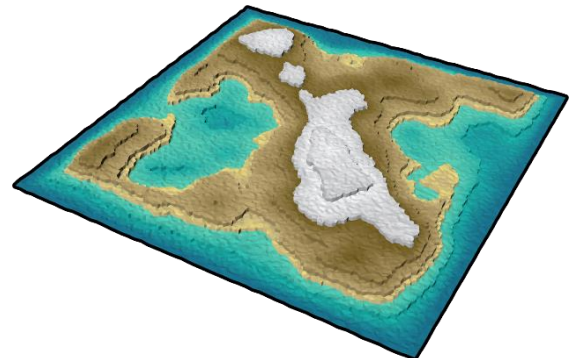

**Figure 18:** *A different world used to introduce new mechanics, enemies, or resources.*

If the algorithm is implemented in a modular way, by using only the vertex's noise value for calculating its Y coordinate, disabling the code responsible for creating terraces, and adding every triangle to the mesh without modifying them makes the terrain look smooth. This is great for games that are not in a top-down view.

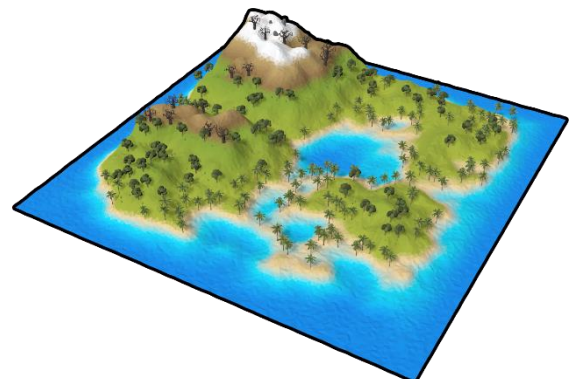

**Figure 19:** *A smooth terrain that can be used in a first of third person game.*

# Evaluation

Evaluating the method involves analysing its quality, performance and practical applications along with some future improvements.

**Quality Assessment**

The algorithm fulfils our initial requirements, generating low-poly terrains with clearly defined biomes separated by terraces and an interesting environment. Although visual appeal is subjective, the algorithm is highly customizable so it can be styled to fit a wide variety of games. It is also recommended to use a top-down view, preferably at an isometric angle and a perspective projection with a smaller field of view for displaying the terrain.

**Performance Analysis**

The performance of the algorithm was evaluated by static code analysis and based on computation time and memory usage across different terrain sizes. Results indicate that smaller terrains generate quickly with minimal resource consumption, making them well-suited for real-time applications. However, as resolution increases, computation time and memory usage grow non-linearly, highlighting the need for optimizations like parallel processing or GPU acceleration.

| Time Complexity | |
| --- | --- |
| Generating Points | $O(n)$ |
| Expanding to Square Shape | $O(n)$ |
| Triangulating Points | $O(n \log n)$ |
| Generating Heightmap | $O(n)$ |
| Forming Terraces | $O(n)$ |
| Colouring Triangles | $O(n)$ |
| Adding Environment | $O(n)$ |

The asymptotic time complexity of the algorithm is $O(n \log n)$, however there are huge constants that are not visible due to the nature of the notation.

For a more precise analysis, terrains of different sizes were generated using an AMD Ryzen 7 3700X 8-Core Processor and an Nvidia GeForce GTX 1660 Ti graphics card. The generations were repeatedly carried out using the Unity Editor, and the measured times were averaged to rule out any edge cases. The table also includes the total memory usage calculated by the built-in C# Garbage Collector.

| Vertex Count | Time | Memory |
| --- | --- | --- |
| ~ 10000 | 22.18 ms | ~ 712 KB |
| ~ 20000 | 48.12 ms | ~ 1408 KB |
| ~ 50000 | 134.77 ms | ~ 3792 KB |
| ~ 100000 | 284.89 ms | ~ 10644 KB |
| ~ 250000 | 781.37 ms | ~ 52560 KB |
| ~ 500000 | 1707.44 ms | ~ 106828 KB |

The asymptotic space complexity of the algorithm is $O(n \log n)$. After plotting the data, it is visible, that the time follows the asymptotic rate, however the memory usage differs from it. Data structures in C# are dynamically allocated, and they usually double their sizes, when their capacity is exceeded. The algorithm uses data structures with this property, which means, after certain thresholds there are big jumps in the memory usage. This is the reason, why the measured memory usage is either smaller or bigger than it is expected.

**Future Improvements**

Based on the results of the analysis, the average computer is more than capable of generating a terrain in a reasonable timeframe, which makes it suitable for computer games. The current implementation of the algorithm runs on a single thread, but it can be parallelized to further decrease its runtime.

Currently, the algorithm is limited to generating a single terrain mesh. By removing the falloff map and squaring the shape of the terrain only once, multiple terrain chunks can be generated and put next to each other, making it scale infinitely. To add further details, the algorithm can be extended to support multiple biome themes within the same world. This can be achieved by generating a biome map using noise functions or a Voronoi partitioning, and sample it while assigning the biomes.

As a final addition, there is no way right now for a simple character to move from one biome level to another. This could be either solved by removing the terraces and making the terrain look like the one on Figure 10, or by selecting random points and their neighbours at the edges of two biome, moving them down and filling the holes they introduce, effectively creating ramps in certain places.

# Conclusion

This paper presented a modern technique for implementing landscape generation with terracing for low-poly computer games. The terrains generated by the algorithm are visually appealing and highly customizable, allowing them to be used in a wide variety of game genres. Even without any further optimization, the runtime of the current implementation is acceptable and suitable to be used in modern games.

Overall, this method provides a solid foundation for procedural low-poly terrain generation, offering efficiency, customization and scalability. Future improvements could make it even more powerful, supporting more complex and interactive environments for game development and other applications requiring procedural world generation.

# References

R. BRIDSON: Fast Poisson Disk Sampling in Arbitrary Dimensions, 2007

S. REBAY: Efficient Unstructured Mesh Generation by Means of Delaunay Triangulation and Bowyer-Watson Algorithm, 1993

K. PERLIN: An Image Synthesizer, 1985

J. OLSEN: Realtime Procedural Terrain Generation, 2004

C. WOLTERING: Triangle.NET. GitHub. [Online]. Available: [link]

S. LAGUE: Sebastian Lague [YouTube Channel]. YouTube. [Online]. Available: [link]

K. LAGUE: Kristin Lague [YouTube Channel]. YouTube. [Online]. Available: [link]

BROKEN VECTOR: Low Poly Tree Pack. Unity Asset Store. [Online]. Available: [link]

JUSTCREATE: Low-Poly Simple Nature Pack. Unity Asset Store. [Online]. Available: [link]